\documentclass[twocolumn,letter]{jpsj3}
\usepackage{txfonts}
\usepackage{color}

\title{Spin-Orbit Coupling in Multilayer Superconductors with Charge Imbalance}

\author{Daisuke MARUYAMA$^{1}$\thanks{E-mail address: marudai@phys.sc.niigata-u.ac.jp}, Manfred SIGRIST$^{2}$, and Youichi YANASE$^{1,3}$ %\\
}

\inst{\address{
$^{1}$Graduate School of Science and Technology, Niigata University, Niigata 950-2181, Japan \\ $^{2}$Theoretische Physik, ETH-Honggerberg, 8093 Zurich, Switzerland \\ $^{3}$Department of Physics, Niigata University, Niigata
950-2181, Japan } %\\
}

\abst{
In this study, we investigate the spin susceptibility in the superconducting state of 
a multilayer system with a layer-dependent Rashba spin-orbit coupling, representing
a locally non-centrosymmetric superconductor. 
We show that the charge imbalance between different layers yields a strong layer
dependence on the susceptibility, most significantly for a weak spin-orbit coupling, 
which is in contrast to a situation with an equally distributed charge. 
These results can be relevant for experimental test in multilayer high-\Tc cuprates as well as in superconducting
CeCoIn$_5$/YbCoIn$_5$ superlattices with more than two layers. 
}

\kword{superconductivity without local inversion symmetry, multilayer superconductor, spin susceptibility}

\newcommand{\Tc}{$T_{\rm c}$ }

\begin{document}

\maketitle

Since the discovery of superconductivity in the heavy-fermion compound without inversion symmetry,
CePt$_3$Si,~\cite{PhysRevLett.92.027003} non-centrosymmetric superconductivity 
has been studied extensively.~\cite{NCSC_Springer} 
Subsequently, several new non-centrosymmetric superconductors with unique properties have been identified among heavy-fermion materials~\cite{JPSJ.73.3129,PhysRevLett.95.247004,JPSJ.75.043703,JPSJ.75.044711} 
and others.~\cite{PhysRevLett.93.247004,JPSJ.76.073708,PhysRevB.74.220502,PhysRevB.76.064527,PhysRevB.76.132508,JPSJ.76.103710}
Many exotic properties induced by antisymmetric spin-orbit coupling 
have been studied in various contexts, such as exotic superconductivity 
and spintronics.~\cite{NCSC_Springer}

On the other hand, many other superconductors lack local inversion symmetry 
while having global inversion symmetry in the crystal structure. 
In contrast to the uniform antisymmetric spin-orbit coupling in non-centrosymmetric systems, 
the antisymmetric spin-orbit coupling appears in such systems with special unit cell structures
with locally losing inversion symmetry.
An interesting class of such ``locally non-centrosymmetric systems'' is the multilayer system
in which the Rashba-type spin-orbit coupling is induced by the local violation of 
mirror symmetry.~\cite{JPSJ.81.034702}
For instance, a recent experiment discovered superconductivity in artificial superlattices 
of the heavy-fermion superconductor CeCoIn$_5$ and the conventional metal YbCoIn$_5$,~\cite{NatPhys.7.849} 
in which superconductivity is induced by multilayers of CeCoIn$_5$. 
The multilayer high-\Tc cuprates represent a further example of superconductors where this type of spin-orbit coupling could play a role.~\cite{PhysRevLett.96.087001,JPSJ.80.043706,JPSJ.81.011008}

We previously investigated the basic properties of multilayer superconductors, such as 
the electron structure, superconducting order parameters, and magnetic properties 
with focus on the effects of Rashba spin-orbit coupling 
arising from the local non-centrosymmetricity. 
It has been shown that Rashba spin-orbit coupling significantly affects superconductivity 
when the spin-orbit coupling is larger than the interlayer single-particle hopping. 
Indeed, an anomalous behavior of the upper critical field has been experimentally observed 
in the superlattice of CeCoIn$_5$, possibly connected with the paramagnetic limiting influenced by spin-orbit coupling.~\cite{PhysRevLett.109.157006} 
Several other superconductors with a locally non-centrosymmetric crystal structure 
have been theoretically investigated.~\cite{JPSJ.79.084701,PhysRevLett.96.127002,PhysRevB.84.184533,
PhysRevB.85.220505,cond-mat.1202.1604,PhysRevB.86.100507} 

In our earlier studies, we focused on the superlattice of CeCoIn$_5$,~\cite{JPSJ.81.034702} 
but neglected the difference in electric potential between 
the inner and outer layers. This approximation may be valid for the superlattices of 
CeCoIn$_5$ since both interlayer hopping and Rashba spin-orbit coupling can be larger than 
the potential difference. On the other hand, it has been shown that 
the charge imbalance due to the potential difference plays an important role in magnetism 
and superconductivity in high-\Tc cuprates with more than two 
layers.~\cite{PhysRevLett.96.087001,JPSJ.80.043706,JPSJ.81.011008,PhysRevLett.94.137003} 
Thus, it is desirable to clarify the effect of spatially inhomogeneous Rashba spin-orbit coupling 
on multilayer superconductors with charge imbalance. 
For this purpose, we study trilayer systems with charge imbalance for simplicity. 

First, we introduce a model Hamiltonian for two-dimensional trilayer superconductors
including spatially modulated Rashba spin-orbit coupling as
{\setlength\arraycolsep{0.2pt}
\begin{eqnarray}
\label{model}
H&=&\sum_{m=1}^3\biggl\{\sum_{\mib{k},s}\varepsilon(\mib{k})c^{\dag}_{\mib{k}sm}c_{\mib{k}sm}
+\sum_{\mib{k},s,s'} \alpha_{m}\mib{g}(\mib{k})\cdot\mib{\sigma}_{ss'}c^{\dag}_{\mib{k}sm}c_{\mib{k}s'm} \nonumber \\
&+&\frac{1}{2}\sum_{\mib{k},s,s'}\Bigl[\Delta_{ss'm}(\mib{k})c^{\dag}_{\mib{k}sm}c^{\dag}_{-\mib{k}s'm}+\rm{h.c.}\Bigr]\biggr\} \nonumber \\
&+&t_{\perp}\sum_{\mib{k},s,\langle m,m'\rangle}c^{\dag}_{\mib{k}sm}c_{\mib{k}sm'}+V\sum_{\mib{k},s}c^{\dag}_{\mib{k}s2}c_{\mib{k}s2},
\end{eqnarray}}
where $c_{\mib{k}sm}$ ($c^{\dag}_{\mib{k}sm}$) is the annihilation
(creation) operator for an electron with a spin $s$ on a layer $m$. 
Two outer layers are denoted $m=1,3$, while the inner layer is denoted $m=2$. 
We consider a simple square lattice and assume the dispersion relation
$\varepsilon(\mib{k})=-2t(\cos{k_{x}}+\cos{k_{y}})-\mu$. 
We choose the unit of energy $t=1$ and fix the chemical potential
$\mu=-1$, which lead to the electron density per site being approximately 0.63 for $V=0$. 
The second term describes the Rashba spin-orbit coupling 
arising from the lack of local inversion symmetry.
The coupling constants $\alpha_{m}$ should be $(\alpha_1,\alpha_2,\alpha_3)=(\alpha,0,-\alpha)$ 
owing to the symmetry. 
We assume a simple Rashba-type form of g-vector 
$\mib{g}(\mib{k})=(-\sin{k_{y}},\sin{k_{x}},0)$ without going into 
microscopic derivation of the antisymmetric spin-orbit coupling.~\cite{JPSJ.77.124711}

The third term introduces an intralayer Cooper pairing via an off-diagonal mean field.
We ignore interlayer pairing as we assume a weak interlayer coupling. 
Owing to the spatially modulated Rashba spin-orbit coupling arising from the 
broken local inversion symmetry, the order parameter $\Delta_{ss'm}(\mib{k})$ involves 
both spin singlet and triplet components. One of the spin singlet and triplet components 
has the same sign between layers, while the other should change its sign. 
For instance, the staggered order parameter of p-wave superconductivity is induced 
by a staggered Rashba spin-orbit coupling when the d-wave Cooper pairing is predominant, 
as in high-\Tc cuprates and CeCoIn$_5$. 
We here neglect this spin triplet component  
because the spin susceptibility discussed below is independent of the 
staggered order parameter, as we showed in ref.~13.
We assume the uniform spin singlet order parameter 
$\Delta_{\uparrow\downarrow m}(\mib{k}) = - \Delta_{\downarrow\uparrow m}(\mib{k}) = 
\psi (\mib{k})$, because the layer dependence on the order parameter 
$\Delta_{\uparrow\downarrow m}(\mib{k})$ hardly affects spin susceptibility.
First, we consider the $s$-wave superconductivity 
[$\psi (\mib{k})= \psi$] for simplicity. 
It will be shown that the model for the $d$-wave superconductivity gives 
qualitatively the same results (see Fig.~\ref{cuprate}).
We take $|\psi| = 0.01$ to be sufficiently small to satisfy the
condition $|\Delta_{ss'm}(\mib{k})|\ll|\alpha_{m}|\ll\varepsilon_{\rm F}$ 
($\varepsilon_{\rm F}$ is the Fermi energy), as realized in most
(locally) non-centrosymmetric superconductors. 

The fourth term in eq.~(1) describes the interlayer hopping of electrons between nearest-neighbor layers.
Since we consider a quasi-two-dimensional system,
we assume an interlayer hopping $t_{\perp} =0.1$ which 
is much smaller than the intralayer hopping $t =1$. In the following 
calculations, we always use this value for $t_{\perp}$, unless stated otherwise. 
The last term of eq.~(1) is the static potential $V$, which induces the charge imbalance between the layers.
We focus on the effect of the potential $V$ in the following part. 
Note that the electron density per site in the inner layer is $n_{\rm in}\simeq 0.78, 0.68, 0.62, 0.56,$ and $0.49$
for $V=-0.5, -0.2, 0, 0.2,$ and $0.5$, respectively, while 
that in the outer layers is $n_{\rm out} \sim 0.63$ independent of $V$.
The total number density slightly changes with $V$ in our calculation. 
We confirmed that the difference in the total number density hardly affects the following results.

\begin{figure}[htbp]
  \begin{center}
    \includegraphics[scale=0.35]{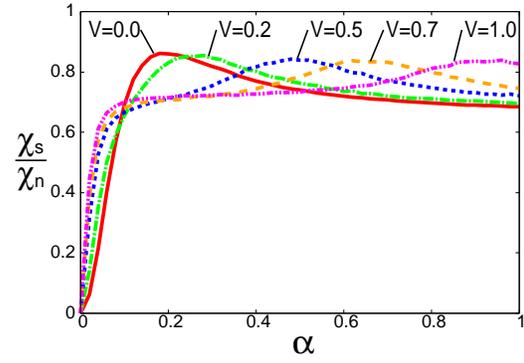}
  \end{center}
  \caption{(Color online)
    Spin susceptibility in the superconducting state at $T=0$ 
    as a function of the spin-orbit coupling $\alpha$. 
    We show the spin susceptibility along the {\it c}-axis normalized by that in the normal state. 
    We assume $V=0$ (solid line), $V=0.2$ (dashed-dotted line), $V=0.5$ (dotted line), 
    $V=0.7$ (dashed line), and $V=1$ (dash two-dotted line), respectively. 
  }
  \label{posus}
\end{figure}

We calculate the spin susceptibility for magnetic fields along the {\it c}-axis. 
The normalized spin susceptibility $\chi_{\rm s}/\chi_{\rm n}$ for various potentials $V$ and 
spin-orbit couplings $\alpha$ is shown in Fig.~\ref{posus}, 
where $\chi_{\rm s}$ and $\chi_{\rm n}$ are the spin susceptibility 
at $T=0$ in the superconducting state and normal state, respectively. 
The normalized spin susceptibility along the {\it ab}-plane is only half of that along the {\it c}-axis, 
as shown in ref.~13.
For $V=0$, our calculation reproduces the results obtained in ref.~13. 
The spin susceptibility vanishes $\chi_{\rm s}/\chi_{\rm n} =0$ for $\alpha =0$
as in the centrosymmetric spin singlet superconductors. 
With increasing $\alpha$, the spin susceptibility first ascends and shows a peak 
around $\alpha \sim t_{\perp}$, 
indicating the interplay between the spin-orbit coupling $\alpha$ and the interlayer hopping $t_\perp$. 
Indeed, this is the fingerprint of crossover from the centrosymmetric superconductivity to 
the non-centrosymmetric superconductivity, which occurs with increasing spin-orbit 
coupling $\alpha$.~\cite{JPSJ.81.034702}  

When we switch on the potential $V$, the peak of the spin susceptibility shifts to $\alpha \sim V$. 
This shift does not mean the suppression of the Rashba spin-orbit coupling. 
Indeed, for a small spin-orbit coupling $\alpha < t_\perp$, the spin susceptibility 
in the superconducting state is significantly enhanced by the potential difference $V$. 
This indicates that the effect of Rashba spin-orbit coupling is enhanced by the 
charge imbalance for reasons discussed below.

\begin{figure}[htbp]
  \begin{center}
    \includegraphics[scale=0.35]{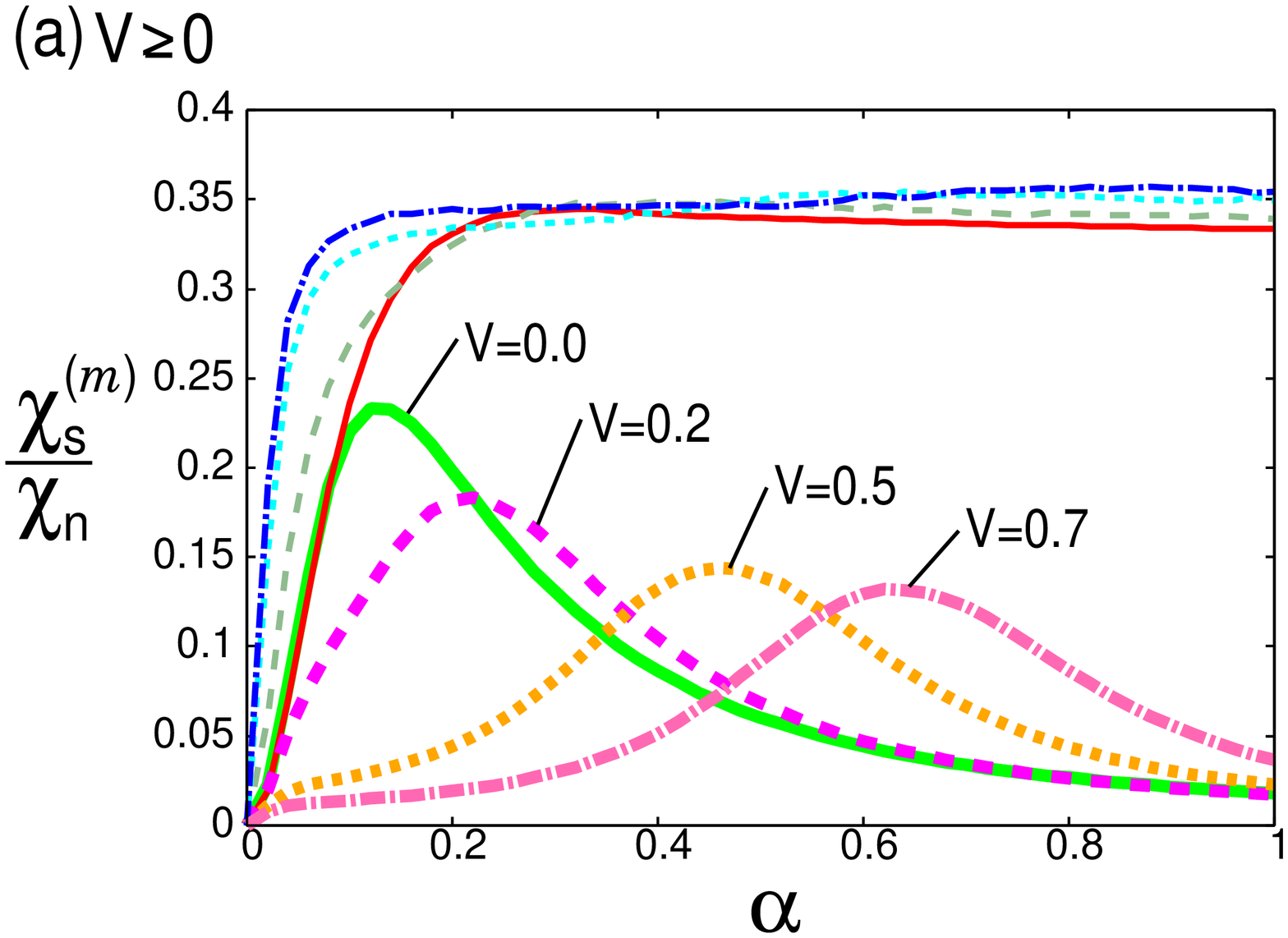}
    \includegraphics[scale=0.35]{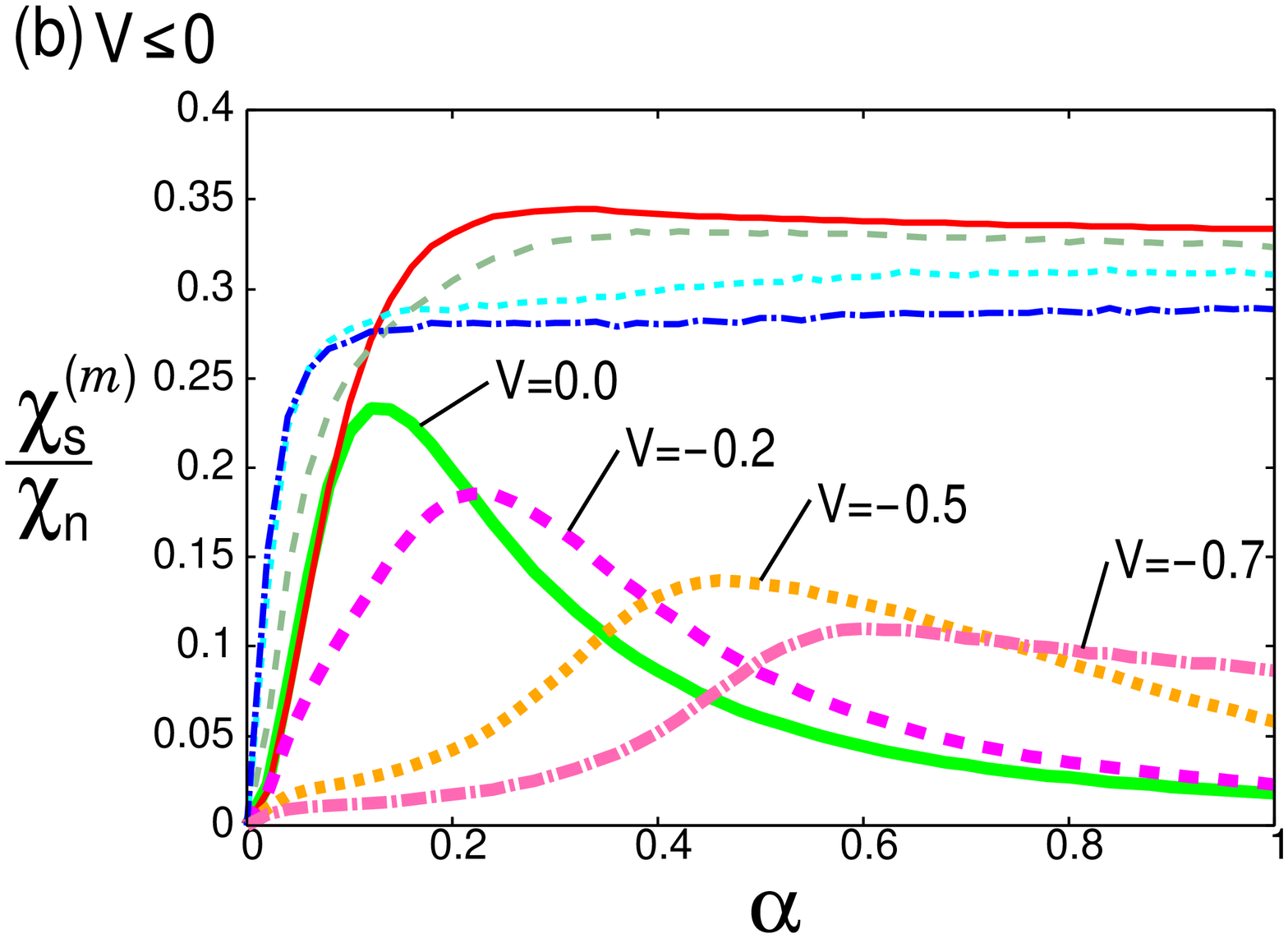}
  \end{center}
  \caption{(Color online)
Contributions of inner and outer layers to spin susceptibility for (a) positive potential $V$ and 
(b) negative potential $V$, respectively. 
The thin and thick lines show the spin susceptibility of the outer layer and inner layer, respectively. 
We show the results for $V=0$ (solid line), $|V|=0.2$ (dashed line), $|V|=0.5$ (dotted line), 
and $|V|=0.7$ (dash-dotted line), respectively. 
  }
  \label{vseihu}
\end{figure}

\begin{figure}[htbp]
  \begin{center}
    \includegraphics[scale=0.35]{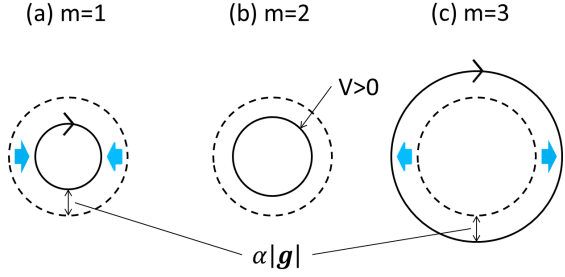}
  \end{center}
  \caption{(Color online)
    Schematic figure of Fermi surfaces in the absence of interlayer hopping $t_{\perp}=0$. 
    We show one of the split Fermi surfaces in the outer layers.  
    The dashed lines show the Fermi surfaces for $\alpha=0$ and $V=0$. 
    The solid lines show the Fermi surface of $m=1,3$ for $\alpha > 0$ and that of $m=2$ for $V > 0$. 
  }
  \label{poband}
\end{figure}

In order to elucidate roles of charge imbalance in more detail, 
we show the contribution of each layer to the spin susceptibility in Fig.~\ref{vseihu}. 
We assume positive and negative potentials $V$ in Figs.~\ref{vseihu}(a) and \ref{vseihu}(b), 
respectively. 
First, we discuss the small spin-orbit coupling $\alpha < t_\perp$. 
We see that the spin susceptibility of the outer layers is enhanced by charge imbalance, 
while that of the inner layer is suppressed. 
Indeed, the spin susceptibility shows a substantial layer dependence for a moderate 
potential difference $V = 0.2$, while the spin 
susceptibilities of the inner and outer layers are nearly identical for $V=0$ and $\alpha < t_\perp$. 
This layer dependence appears through the decoupling of the inner and outer layers due to charge imbalance. 
As the potential difference $|V|$ increases, the interlayer hybridization due to $t_\perp$ 
is suppressed. Then, the effect of Rashba spin-orbit coupling is enhanced; therefore, 
the spin susceptibility of the outer layers increases, while that of the inner layer decreases.  
When the layers are completely decoupled from each other, the spin susceptibility of the outer layers is not 
suppressed by the superconductivity as in uniformly non-centrosymmetric superconductors,~\cite{NewJPhys.6.115}  
while that of the inner layer is completely suppressed as in the centrosymmetric superconductors. 
The total spin susceptibility shown in Fig.~\ref{posus} is enhanced by the potential difference $|V|$ 
since the contribution of the outer layers is larger than that of the inner layer. 

Next, we discuss the large spin-orbit coupling $\alpha > t_\perp$. 
We see the peak of the spin susceptibility around $\alpha \sim |V|$ in Fig.~\ref{posus}. 
This peak indicates the crossover in the electronic state. 
In order to illustrate this crossover, we show the schematic figures of Fermi surfaces for 
$t_\perp =0$ (see Fig.~\ref{poband}). 
The Fermi surfaces are identical for all layers at $V=\alpha=0$ as shown by the dashed lines 
in Fig.~\ref{poband}. 
When the potential of the inner layer $V >0$ is turned on, the Fermi surface of the inner layer shrinks, 
as shown by the solid line in Fig.~\ref{poband}(b). 
Then, the mismatch of Fermi surfaces between $m=1$ and $m=2$ and between $m=2$ and $m=3$ 
suppresses the interlayer hybridization. 
When we switch on the Rashba spin-orbit coupling, the Fermi surfaces of the outer layers are split. 
One of the split Fermi surfaces having a positive spin 
helicity is drawn by the solid lines in Figs.~\ref{poband}(a) and \ref{poband}(c).  
The Fermi surface of the $m=1$ layer shrinks, while that of the $m=3$ layer is enlarged, because the spin-orbit couplings 
have opposite signs, i.e., $\alpha_1 = -\alpha_3$. 
Then, the mismatch of Fermi surfaces between $m=1$ and $m=2$ is removed, 
and therefore, the quasiparticle strongly hybridizes between the inner layer $m=2$ and the outer layer $m=1$. 
This hybridization enhances the spin susceptibility of the inner layer without decreasing the 
spin susceptibility of the outer layers, as shown in Fig.~2. This is the reason why 
the total spin susceptibility shows a peak around $\alpha \sim |V|$. 
When the spin-orbit coupling $\alpha$ is further increased, the interlayer hybridization is 
suppressed again, and then the spin susceptibility of the inner layer is decreased. 

\begin{figure}[htbp]
  \begin{center}
    \includegraphics[scale=0.35]{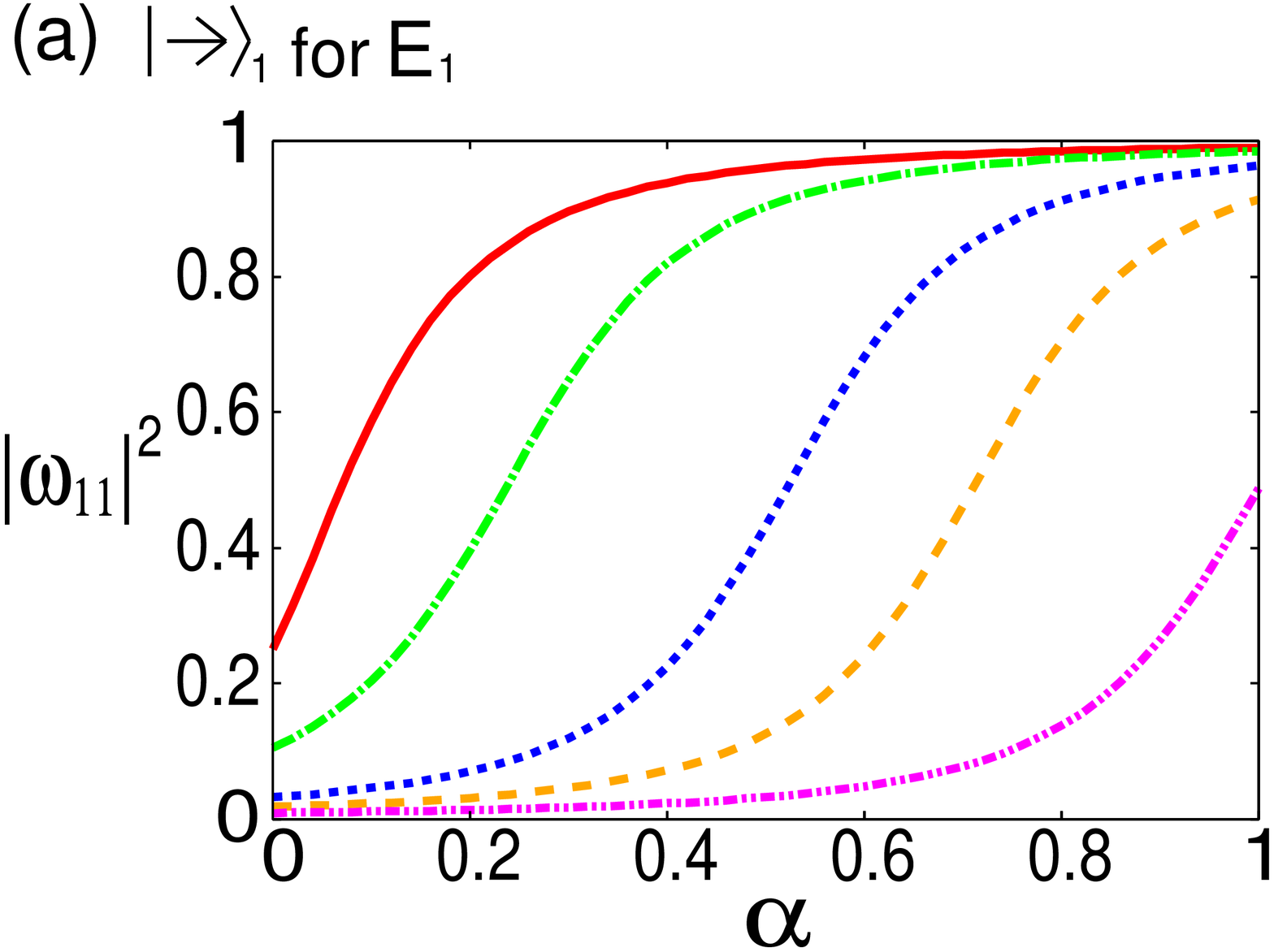}
    \includegraphics[scale=0.35]{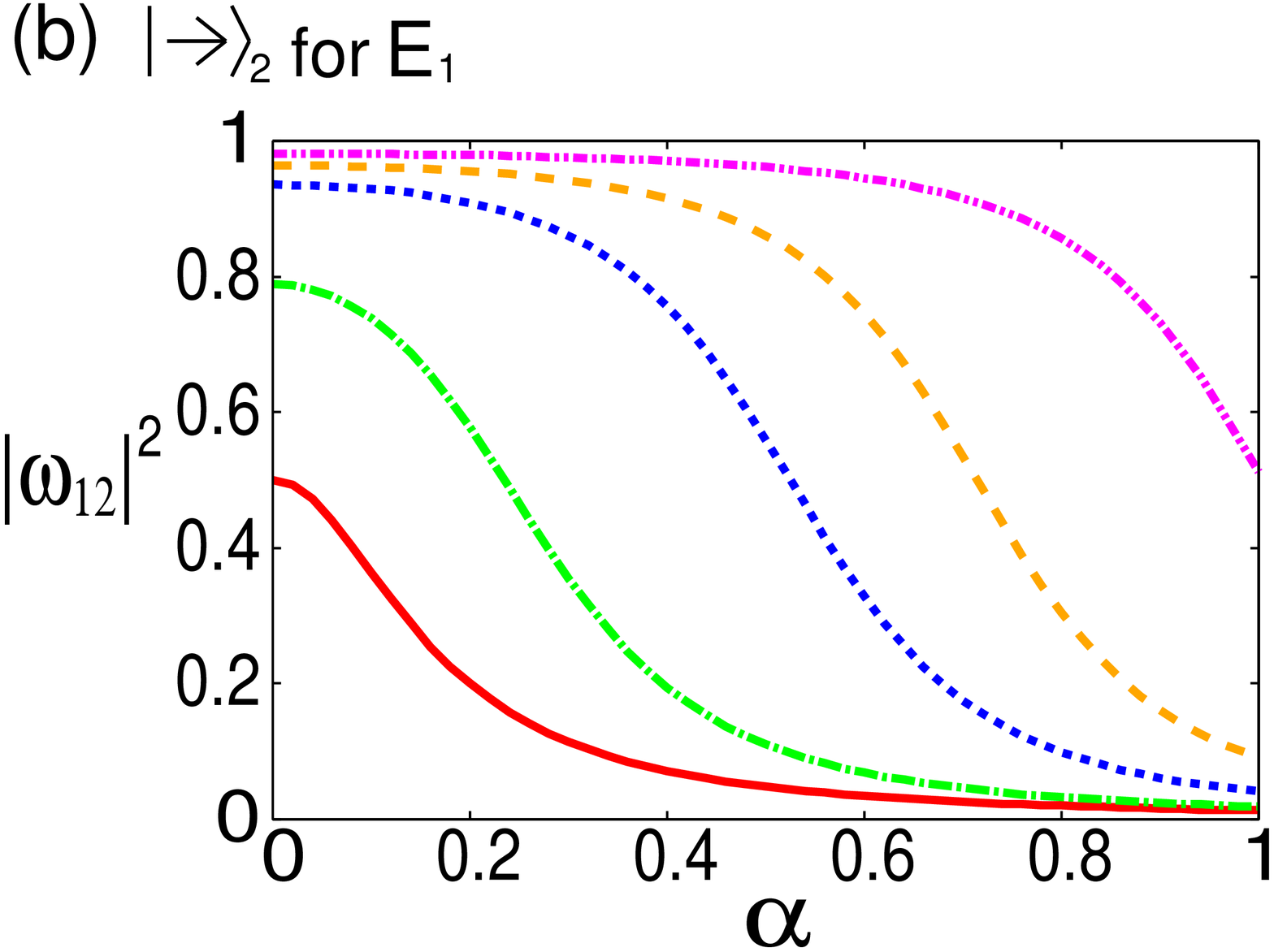}
    \includegraphics[scale=0.35]{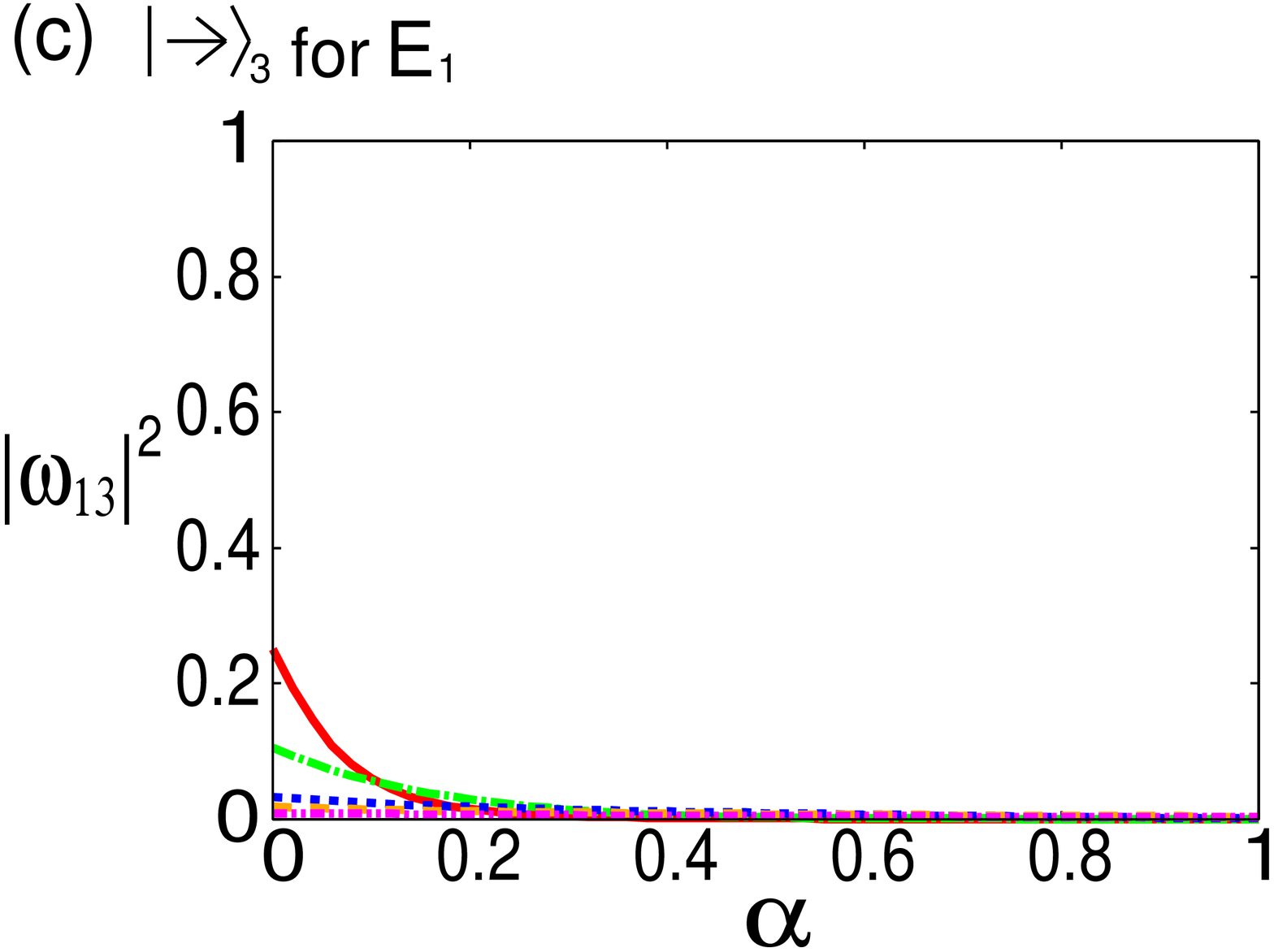}
  \end{center}
  \caption{(Color online)
    Single-particle weight $|\omega_{jm}|^2$ of the band $j=1$ on the layer (a) $m=1$, (b) $m=2$, and (c) $m=3$. 
    We assume the following potentials on the inner layer: 
    $V=0$ (solid lines), $V=0.2$ (dash-dotted lines), $V=0.5$ (dotted lines), $V=0.7$ (dashed lines), 
    and $V=1$ (dashed two-dotted lines).}
  \label{expro}
\end{figure}

\begin{figure}[htbp]
  \begin{center}
    \includegraphics[scale=0.35]{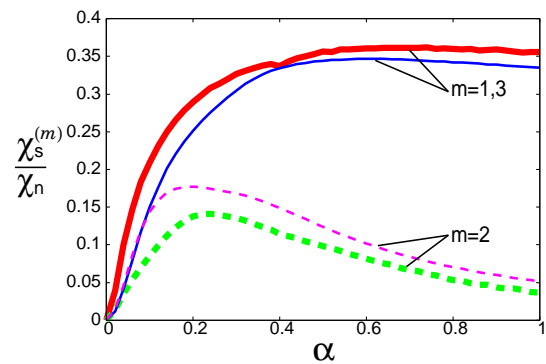}
  \end{center}
  \caption{(Color online)
    Spin susceptibility in model for $d_{x^2-y^2}$ pairing state in multilayer high-\Tc cuprates. 
    We show the spin susceptibility of the inner (thick dashed line) and outer (thick solid line) layers 
    for $V=-0.2$. 
    The thin lines show the results for $V=0$. 
    The other parameters are described in the text. 
  }
  \label{cuprate}
\end{figure}

In order to substantiate the above picture, we now show the single-particle wave function in the normal state. 
We focus on the single-particle state with momentum $\mib{k}=(0,k_{\rm F})$ and take 
the {\it x}-axis for the quantization axis of spins, as in ref.~13. 
Then, the single-particle Hamiltonian is block-diagonalized with the block
{\setlength\arraycolsep{0.5pt}
\begin{eqnarray}
&& \hspace*{-3mm}
\left(
\begin{array}{ccc}
\varepsilon(\mib{k})+\alpha|\mib{g}(\mib{k})| & t_{\perp} & 0 \\
t_{\perp} & \varepsilon(\mib{k})+V & t_{\perp} \\
0 & t_{\perp } & \varepsilon(\mib{k})-\alpha|\mib{g}(\mib{k})| \\
\end{array}
\right)
\hspace*{2mm}
{\rm for }
\hspace*{2mm}
\left(
\begin{array}{ccc}
|\rightarrow\rangle_{1}\\
|\rightarrow\rangle_{2}\\
|\rightarrow\rangle_{3}\\
\end{array}
\right),
\label{mat33}
\end{eqnarray}
}
in the subspace of states describing the electrons on the $m$ layer with a spin pointing in the positive $x$-direction. 
Diagonalizing the $3\times3$ matrix, we obtain three bands with energy 
$E_1(\mib{k}) > E_2(\mib{k}) > E_3(\mib{k})$ for the given $k$-vector and the wave function 
$|j,\rightarrow\rangle = \sum_m \omega_{jm} |\rightarrow\rangle_{m}$, where $ j $ is the band index. 
Figure~\ref{expro} shows the representative weights $ |\omega_{jm} |^2 $ for the band $j=$1 
on the three $m$ layers. For a vanishing spin-orbit coupling ($\alpha=0$), the single-particle weight is localized
on one to the layers for a finite $V$, i.e., on the inner $m=2$ layer in the case of band $j=1$. As the
spin-orbit coupling is turned on, the layers begin to hybridize, leading to the strongest hybridization for $\alpha \sim V$
(largest change in weight). Eventually, at a larger $ \alpha $, the weight shifted to the $ m =1 $ layer. 
Most notably, for the strongest hybridization between the inner and outer layers, the effect of spin-orbit coupling is also transferred
to the inner layer giving rise to the maximum susceptibility. 

Charge imbalance may pave the way for observing the  ''local non-centrosymmetricity'' in multilayer superconductors
such as high-\Tc cuprates even for a comparatively weak spin-orbit coupling by studying the local spin susceptibility.  
The local spin susceptibility is, in principle, accessible by site-selective NMR experiments through
Knight shift measurements.~\cite{PhysRevLett.96.087001,JPSJ.80.043706,JPSJ.81.011008} 
Experimental investigations have not focused on the aspect of the local non-centrosymmetricity so far. In order to motivate
such studies, we consider here a simple tight-binding model as an approximation of the generic band structure 
of trilayer high-\Tc cuprates in the overdoped regime in order to avoid complications with an antiferromagnetic order.
In-plane hopping involves nearest- and next-nearest-neighbor hoppings,
leading to $\varepsilon(\mib{k})=-2t_1(\cos{k_x}+\cos{k_y})+4t_2\cos{k_x}\cos{k_y}-\mu$,
and the interlayer hopping $t_{\perp}$ is simply the nearest neighbor.
We introduce uniform $d_{x^2-y^2}$ order parameter 
$\psi (\mib{k})= \psi (\cos k_{\rm x} - \cos k_{\rm y})$ for high-\Tc cuprates and 
superlattice CeCoIn$_5$/YbCoIn$_5$.
Experimental studies give for such systems a charge distribution
among the layers of $n_{\rm{in}}\simeq0.9$ and $n_{\rm{out}}\simeq0.8$.~\cite{JPSJ.81.011008}
These values are obtained by choosing the parameters $(t_1,t_2,t_\perp,\mu,V)=(1,0.3,0.1,-1.08,-0.2)$, which we now use to
calculate the spin susceptibility in the superconducting phase for various spin-orbit couplings $\alpha$, as shown in Fig.~\ref{cuprate}. 
We find a clear difference in the spin susceptibility between the inner and outer layers 
even for a small spin-orbit coupling $\alpha < 0.04$. This is in contrast to the result for $V=0$. 
Thus, the moderate charge imbalance in our model for multilayer cuprates significantly enhances 
the effect of the Rashba spin-orbit coupling on the outer layers. 
Naturally, the Rashba spin-orbit coupling would also play a role in the antiferromagnetic and 
superconducting states of multilayer cuprates. In particular, the difference in the magnetic response between the inner and outer 
layers affects the superconducting phase structure in a magnetic field. 
For example, the pair-density wave state that has been discussed recently~\cite{PhysRevB.86.134514} would be stabilized 
by the reduced interlayer hybridization due to the cooperation of spin-orbit coupling and 
charge imbalance. A more detailed discussion of the interplay of magnetism and superconductivity in high-\Tc cuprates in the light of the
local non-centrosymmetricity will be given elsewhere. 

In summary, we have investigated the effect of charge imbalance on the spin susceptibility of 
trilayer superconductors by taking into account the Rashba spin-orbit coupling arising from the 
local non-centrosymmetricity. As an important result, we find that charge imbalance enhances
the spin susceptibility of the outer layers, because the interlayer hybridization is suppressed by the mismatch of Fermi surfaces. 
The same effect leads to a reduction in the spin susceptibility of the inner layer. 
For a small spin-orbit coupling and a moderate charge imbalance, spin polarization 
mainly occurs in the outer layers, thus showing that the effect of spin-orbit coupling is enhanced 
by charge imbalance. This may also be the case for multilayer high-\Tc cuprates. 
When spin-orbit coupling is increased, a strong hybridization occurs 
between the inner and outer layers around $\alpha \sim |V|$, and then, 
the spin susceptibility of the inner layer increases substantially. 
These synergistic and/or competing roles of Rashba spin-orbit coupling and charge imbalance can be 
tested by experiments on multilayer superconductors such as high-\Tc cuprates as well as on the 
superlattice of CeCoIn$_5$/YbCoIn$_5$. 

\section*{Acknowledgements}
The authors are grateful to S. K. Goh, H. Shishido, T. Shibauchi, Y. Matsuda, 
M. H. Fischer, and D. F. Agterberg
for fruitful discussions. 
This work was supported by a Grant-in-Aid for Scientific Research 
on Innovative Areas ``Heavy Electrons'' (No. 23102709) from MEXT, 
Japan. It was also supported by a Grant-in-Aid for Young Scientists 
(B) (No. 24740230) from JSPS. 
We are also grateful for the financial support from the Swiss Nationalfonds, the NCCR MaNEP, 
and the Pauli Center of ETH Zurich. 

\renewcommand{\refname}{}

\end{document}